# Study of waveguide background at visible wavelengths for on-chip nanoscopy

**DAVID A. COUCHERON,**[1] **ØYSTEIN I. HELLE,**[2] **JAMES S. WILKINSON**[3]**, GANAPATHY SENTHIL MURUGAN**[3]**, CARLOS DOMÍNGUEZ**[4]**, HALLVARD ANGELSKÅR**[5] **AND BALPREET S. AHLUWALIA**[1]

[1]*UiT The Arctic University of Norway, Klokkargårdsbakken 35, 9037 Tromsø, Norway*
[2]*Chip NanoImaging AS, Sykehusvegen 23, 9019, Tromsø, Norway*
[3]*Optoelectronics Research Centre, University of Southampton, Southampton SO17 1BJ, U.K.*
[4] *Instituto de Microelectrónica de Barceloana (IMB-CNM, CSIC), Campus UAB, 081930-Bellaterra, Spain*
[5] *SINTEF Digital, Dept. of Microsystems and Nanotechnology, Gaustadalleen 23C, 0373 Oslo, Norway*
*[\*balpreet.singh.ahluwalia@uit.no](*balpreet.singh.ahluwalia@uit.no)*

**Abstract:** On-chip super-resolution optical microscopy is an emerging field relying on waveguide excitation with visible light. Here, we investigate two commonly used high-refractive index waveguide platforms, tantalum pentoxide ($Ta_2O_5$) and silicon nitride ($Si_3N_4$), with respect to their background with excitation in the range 488-640 nm. The background strength from these waveguides were estimated by imaging fluorescent beads. The spectral dependence of the background from these waveguide platforms was also measured. For 640 nm wavelength excitation both the materials had a weak background, but the background increases progressively for shorter wavelengths for $Si_3N_4$. We further explored the effect of the waveguide background on localization precision of single molecule localization for *direct* stochastic optical reconstruction microscopy (*d*STORM). An increase in background for $Si_3N_4$ at 488 nm is shown to reduce the localization precision and thus the resolution of the reconstructed images. The localization precision at 640nm was very similar for both the materials. Thus, for shorter wavelength applications $Ta_2O_5$ is preferable. Reducing the background from $Si_3N_4$ at shorter wavelengths via improved fabrication will be worth pursuing.

## 1. Introduction

Integrated photonics is a rapidly expanding field of research. It started out with integration of optical devices onto 2D substrates for telecommunications [1], but soon expanded to much more complex components. Today, both passive components such as multi-mode interference devices [2], as well as active components such as lasers [3] have been implemented. This has greatly increased the usefulness of the technology and now encompasses not only telecommunications, but diverse fields such as quantum mechanics [4] and optical trapping [5,6]. An important biological application has been waveguide-based biosensors [7]. A wide range of applications have been shown, ranging from Raman spectroscopy [8–10] to fluorescence imaging [11]. For sensing applications, integrated photonics can offer signal enhancement [8], size reductions [12] and many other benefits. High refractive index materials are of particular interest as they offer high sensitivity in smaller footprints [13]. The choice of waveguide material is based on several criteria, such as the wavelength of interest. Traditionally, integrated photonics have been developed for near infra-red wavelength for telecommunications applications. However, the increased interest in bio-applications is pushing the field closer to and into the visible range. For visible wavelengths, different material platforms have been investigated, with popular platforms being silicon nitride ($Si_3N_4$) [8,14], tantalum pentoxide ($Ta_2O_5$) [10,15] and titanium dioxide ($TiO_2$) [9].

Development of Raman spectroscopy using photonic chips has seen a surge of interest over recent years [8,9,15–18]. Different waveguide designs [5,16,19] and collection methods [5,20] have been suggested. Highly efficient waveguides based Raman sensors have been made, rivalling the sensitivity of the traditional confocal Raman microscope [21]. Dhakal et al. have recently measured biological sub-monolayers using a spiral waveguide design [8]. Most of the chip-based Raman spectroscopy work until now has been performed using a 785 nm laser. Lately, there has been a growing interest in visible light for Raman spectroscopy [10,15,22]. A lower excitation wavelength increases the signal strength, as Raman scattering scales with $\lambda^{-4}$. Furthermore, visible wavelengths are also demonstrated to be useful for resonance Raman spectroscopy, such as 532 nm for hemoglobin [10,23,24]. The shorter wavelength can, however, also increase any background from fluorescence.

Recently, fluorescence based super-resolution optical microscopy using photonic integrated circuits was introduced [25], which exclusively uses visible wavelengths. Until now, three different well-established super-resolution methodologies have been demonstrated using a waveguide platform. These are: 1) single molecule localization microscopy (SMLM) such as direct stochastic optical reconstruction microscopy (*d*STORM) [25,26], 2) structured illumination microscopy (SIM) [27] and 3) fluorescence intensity fluctuations based methods, such as on-chip super-resolution radial fluctuation microscopy (SRRF) [28], on-chip super-resolution optical fluctuation imaging (SOFI) [29] and on-chip entropy-based super-resolution imaging (ESI) [25]. Among the existing methods, on-chip SMLM has provided the highest resolution. The SMLM techniques rely on precise localization of individual fluorescent probes [30–34]. In traditional fluorescence imaging, the entire sample is labelled with a fluorescent dye that is excited by a laser and emits red-shifted light, which can then be collected to create an image of the sample. The imaging process introduces a fundamental resolution limit known as the diffraction limit. SMLM techniques can surpass this limit. This is achieved by imaging subsets of fluorophores that are well separated and fitting the data in order to determine the localization precisely. A peak fitting procedure is then used to localize all the different fluorophores at very high precision, and from the localizations a super-resolved image is generated. Super-resolution imaging is often performed using evanescent field excitation from total internal reflection objective lenses. This removes most of the out of focus background and thus gives an improved result. In 2017 Diekmann et al. performed super-resolution imaging using optical waveguides for excitation [25]. The new approach offered miniaturization and high throughput imaging. A major advantage is the separation of the excitation and collection pathways. The evanescent field excitation is provided by the waveguide, removing any restrictions on collection optics. As such, arbitrarily large fields-of-view can be imaged [35]. The platform is also live-cell compatible and the same platform can be used for other applications such as Raman spectroscopy [10] or optical trapping [6]. Thus, several research fields are looking towards visible light applications with integrated photonics, and to this end performing systematic investigation of the background at visible lights for these waveguide platforms is useful.

Whenever light is guided in a waveguide it will also interact with the waveguide platform. Thus, any background from the waveguide material will have implications on the performance for on-chip spectroscopy and on-chip nanoscopy. Two important origins for background in these applications are Raman scattering and background fluorescence from the waveguide platform itself. Raman scattering will yield the same spectrum (frequency shift) independent of excitation wavelength, except for a scaling in intensity. Fluorescence, however, is strongly dependent on excitation wavelength. Any background, be it Raman or fluorescence based, can cause problems for Raman spectroscopy applications. The particular implications of a background depend on the actual spectrum – there might be applications where a Raman background can be filtered out, whereas a broadband fluorescence background is hard to eliminate. For imaging applications, the spectral distribution does not matter as the camera sensor only measures the total intensity. As fluorescent samples usually emit very strongly, imaging applications should tolerate a background better than spectroscopic applications. However, in case of SMLM the signals come from single fluorophore molecules and therefore the number of photons reaching the camera is usually low. Thus, any background from the waveguide platform will reduce the localization precision as the data fitting will be done over a higher base level [36]. Furthermore, an inhomogeneous background is particularly challenging, as it can lead to a bias in localizations where the localization precision is different in the two lateral dimensions. Fluorescence emission is stronger when pumped in the visible range. As such, the signal-to-background (S/B) ratio for Raman spectroscopy might fall despite the increased Raman scattering at shorter wavelengths.

As interest in waveguide-based optical nanoscopy and Raman spectroscopy keeps growing, it is important to investigate the most relevant platforms for these applications and how any potential background will affect the measurement results. Fluorescence-based nanoscopy is performed solely in the visible range and is thus potentially subject to strong background fluorescence from the waveguide platform. In this work, we have investigated $Ta_2O_5$ from one fabrication procedure and $Si_3N_4$ from two different fabrication procedures as potential candidates for short visible wavelength imaging, as well as for use in Raman spectroscopy. The background is first assessed by imaging fluorescent beads to obtain an estimate of signal-to-background ratio for the different platforms at 640 nm, 561 nm and 488 nm excitation. The spectra for the $Si_3N_4$ with strongest background was measured to investigate the origin. Finally, we investigated the effect of a background on *d*STORM reconstruction by performing blinking experiments at 640 nm and 488 nm. The results show a noticeable difference in background both between platforms and with varying excitation wavelength.

## 2. Methods

### 2.1 Experimental set-up

The waveguide chips were mounted on a vacuum chuck and light was coupled into them using a 50x/0.5 NA coupling lens. A PDMS frame was put on top of the waveguide region of interest to contain the imaging medium and sealed off with a cover slip. We used an upright microscope for imaging. The chip set-up is illustrated in Fig. 1a.

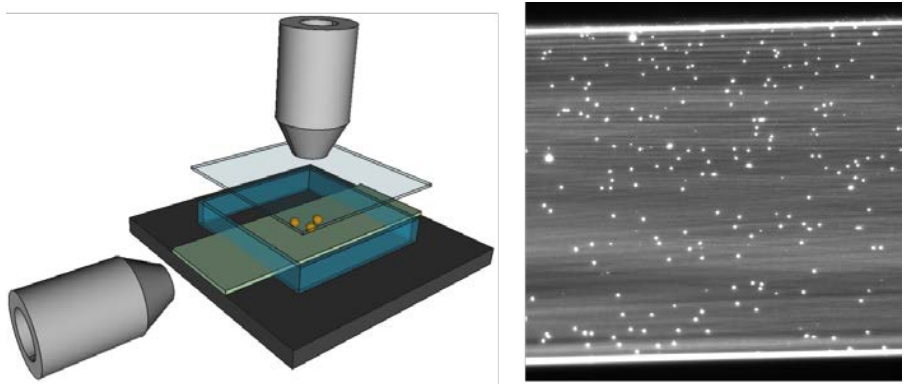

Figure 1: (a) Schematic of the waveguide structure with beads on top. The sample is contained in a PDMS frame with imaging medium and sealed off with a cover slip. The horizontal coupling lens is used to couple light into the waveguide of interest. (b) An example of TIRF imaging of fluorescent beads with a silicon nitride waveguide with a relatively strong background. The multi-mode pattern is clearly visible.

### 2.2 Waveguide fabrication

Three different waveguide platforms were used in this work. The $Ta_2O_5$ platform was fabricated at the Optoelectronics Research Centre (ORC), University of Southampton. Two different $Si_3N_4$ platforms were used in this work: 1) $Si_3N_4$ fabricated at the Institute of Microelectronics Barcelona termed $Si_3N_4$-B and 2) $Si_3N_4$ fabricated by SINTEF, MiniLab, Oslo termed $Si_3N_4$-S.

The optimization of fabrication procedures for $Ta_2O_5$ waveguides can be found in previous literature [37]. A thin layer (150 to 220 nm) of $Ta_2O_5$ was sputtered onto an oxidized silicon substrate ($SiO_2$ thickness ~ 5 µm). Magnetron assisted RF sputtering was employed and the deposition was optimized using an iterative process. The optimized parameters were substrate temperature, magnetron power and the oxygen flow rate. The optimized parameters chosen were 200°C substrate temperature, magnetron power of 300 W, chamber pressure of 10 mTorr and $O_2$ and Ar flow rates of 5 and 20 sccm, respectively. Post-deposition, standard photolithography was used to pattern a photoresist mask with straight channels and ion-beam milling was used to fabricate fully etched channel waveguides. The sidewall roughness, which is the major source of propagation loss, was reduced by optimization of the ion-beam milling process. Finally, plasma ashing was performed to remove the remaining photoresist after the etching process. Post-fabrication, the waveguides were annealed for 5 hours at 600°C to reduce the propagation loss.

Various silicon nitride technologies for photonic applications have spread in the past decade by means of foundry models that mirror electronic integrated circuit industry developments of the past century [38]. Many of the devices and circuits developed with this technology have been employed in biomedical and chemical sensing applications within the visible wavelength range [39]. The $Si_3N_4$-B waveguides were fabricated first by thermally growing a silicon dioxide layer with a thickness of 2 µm on silicon wafers. The thin $Si_3N_4$ layer (150 nm thick) was deposited using low-pressure chemical vapor deposition (LPCVD) at 800°C. Conventional photolithography processes were followed to deposit photoresist, UV-exposure and for defining the waveguide geometry. Photolithography was followed by reactive ion etching (RIE) to fabricate the waveguides. The remaining photoresist was then removed, and a top cladding layer was deposited by plasma-enhanced chemical vapor deposition (PECVD) at 300°C of thickness 1.5 µm. At the imaging areas, the top cladding was opened using RIE and wet etching to enable seeding of the cells. Further details of the fabrication process for $Si_3N_4$ waveguides can be found elsewhere [40].

The $Si_3N_4$-S waveguides were fabricated by first growing a thermal oxide of 2.4 µm on silicon wafers. A thin $Si_3N_4$-layer of 140 nm thickness was deposited by low-pressure chemical vapour deposition (LPCVD) at 770°C. These samples were studied further as slab waveguides.

### 2.3 Imaging of fluorescent beads

The background intensity was first investigated by imaging a sparse sample of fluorescent beads using all three platforms (one for $Ta_2O_5$ and two for $Si_3N_4$). 500 µm wide strip waveguides were used for $Ta_2O_5$ and $Si_3N_4$-B whereas a slab waveguide was used for the $Si_3N_4$-S. The waveguides were treated for 30 s in a plasma cleaner to make the surface more hydrophobic, then 0.5 µL of 100 nm TetraSpeck beads was added and allowed to dry. Once the sample was dried, DI water was added to fill the PDMS chamber completely and sealed it off using a cover glass. Since the waveguides were multi-moded, approximately 1000 images were captured for each wavelength while looping the coupling objective along the coupling facet to reduce the mode pattern. For the slab waveguide, only a single image was used. All platforms were imaged with 640 nm, 561 nm and 488 nm excitation sequentially, through a matching band pass and long pass filter that would be used for fluorescence imaging.

A local thresholding algorithm (Phansalkar [41]) in ImageJ image processing software was used to identify all beads in each image. The average maximum beads intensity was measured. The maximum intensity of each bead was used in order to reduce the measurement uncertainty by averaging varying sizes of beads. An average background intensity was measured by excluding the localized beads with a 7 pixel padding.

### 2.3 Measuring background spectra

The background spectrum of the $Si_3N_4$-B was measured for excitation at 640 nm, 561 nm and 488 nm. The light was coupled into 500 μm wide strip waveguides using a 50x/0.5NA objective lens. Spectroscopic measurements were done by coupling the guided light out of the end of the waveguide and passing it through a suitable long-pass filter to remove the excitation light. The filtered signal was then coupled into a multi-mode optical fiber and passed into a Yokogawa spectrometer (AQ6373B). The spectra were collected using a collection time of 30 s and averaged over 5 measurements.

*2.4 dSTORM imaging and reconstruction*

The effect of the background emission on *d*STORM imaging was measured by investigating fluorophore monolayers. The monolayers were prepared on clean waveguides inside an approximately 170 um thick PDMS frame. The PDMS frame was filled with 0.01% Poly-L-Lysine (PLL) and incubated for 10 minutes. All the PLL was then drained from the chamber and the chamber was rinsed with deionized water. AF-647-dextran conjugate was then added at a concentration of 0.2 μg/ml. For imaging at 488 nm AF-488-dextran conjugate was added instead. Imaging was performed with 70 mW power from the laser coupled into the waveguides and with 30 ms exposure time. We used the following blinking buffer: 22.5 μL PBS, 20.0 μL glucose, 5 μL MEA and 2.5 μL enzymes. The measurements were performed both on the $Si_3N_4$-B and $Ta_2O_5$ waveguides.

In single molecule localization microscopy, sometimes it is desirable to simultaneously excite with a shorter wavelength, e.g. 488 nm even when imaging is performed with longer excitation wavelength, 647 nm. Activation with shorter wavelength assists to increase the blinking events and to get more fluorophores back into a fluorescing state. We investigated the effect of adding 488 nm excitation while imaging with 640 nm. Monolayers were prepared using AF-647 on the $Si_3N_4$-B waveguides. Half of the dataset was acquired using only 640 nm and for the latter half we added 488 nm excitation at 10% of the 640 nm power.

We also performed simulations to determine whether the multi-mode pattern of the background introduces any localization bias to the reconstruction. First, we generated a density filter for simulating realistic blinking data in ImageJ with a patterned excitation light. This was done by loading an image of the multi-mode pattern without any beads into ImageJ. The camera offset was subtracted, since parts of the waveguide will have zero intensity and thus no blinks. A Gaussian blur was then applied. The resulting image was used as the density filter. Simulated data stacks of 100 images were generated in the Thunderstorm plugin with both a low blink intensity distribution (200-500 photons) and a high (1200-1500 photons) with zero camera offset. The background pattern image was then added to the stack to give a realistic background pattern and then finally we reconstructed the stacks. Using the built-in evaluation function on the reconstructed data and the ground truth data from the simulation, we got a root mean square error (RMSE) in both the vertical and horizontal direction.

Thunderstorm [42], a plugin for ImageJ, was used for all *d*STORM image reconstruction. The datasets were filtered to remove hot pixels.

## 3. Results

*3.1 Background*

The first measurement performed was imaging of TetraSpeck beads in order to gauge the background intensity. The resulting images are presented in Figure 2. Representative waveguides were chosen from a batch and checked for damage and contaminants. The multi-mode interference patterns are clearly visible in $Si_3N_4$ for shorter wavelengths. $Si_3N_4$-S exhibits strong lines along the waveguide, which is expected as it is a slab waveguide. $Si_3N_4$-B on the other hand has a typical multi-mode pattern. In order to quantify the background intensity, we compared the average maximum bead intensity to the average background intensity. The beads were localized using a local thresholding algorithm, due to the inhomogeneous background in the slab waveguides. The average bead intensity can be used to gauge the guided power inside the waveguide, so all background measurements were adjusted for differences in guided intensity between the platforms for the same wavelength. All measurements were adjusted were to the bead intensity of $Ta_2O_5$. The resulting measurements are presented in Table 1. It is, however, not straightforward to compare bead intensities at different excitation wavelengths as several factors that determine bead signal strength depend on wavelength, such as waveguide confinement factor and quantum yield of the beads. $Ta_2O_5$ has no detectable background besides the camera offset (approximately 100 counts). The background for $Si_3N_4$ is larger than for $Ta_2O_5$ at all wavelengths, although by a very small amount at 640 nm excitation. For shorter wavelengths, $Si_3N_4$ is increasingly challenging due to increasing background. The disparity between the platforms at all excitation wavelengths, and especially at 488nm is important to note.

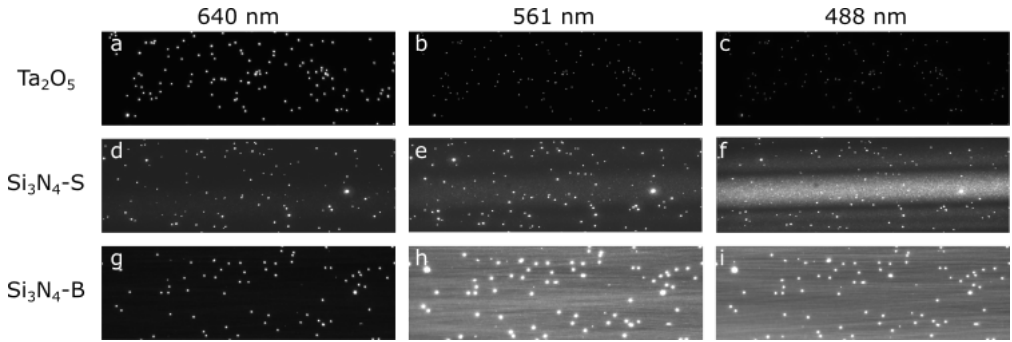

Fig. 2. Fluorescent beads imaged with waveguide excitation using different waveguide platforms with different excitation wavelengths. (a-c) Beads on $Ta_2O_5$ waveguides, (d-f) on $Si_3N_4$-S and (g-i) on $Si_3N_4$-B. Panels e and f exhibit a region of stronger background and this is because it is a slab waveguide and only a single image was acquired. All images are approximately 15 μm by 50 μm.

Table 1: Overview of background from the different waveguides with different excitation wavelengths. Fluorescent beads were used to adjust for similar guided power for the different waveguides at the same excitation wavelength. The backgrounds given are adjusted for similar beads strength for each wavelength. The camera used for the experiments has an offset of 100 counts. BG = background.

| Wavelength (nm) | Average beads signal (counts) | $Ta_2O_5$ BG (counts) | $Si_3N_4$-S BG (counts) | $Si_3N_4$-B BG (counts) |
|---|---|---|---|---|
| 640 | 5132 | 100* | 102 | 107 |
| 561 | 1478 | 100* | 107 | 127 |
| 488 | 2711 | 100* | 150 | 202 |

It is imperative to understand the spectrum of the detected background: A background that is not spectrally overlapping with the sample could easily be filtered out with the correct filter set. We have previously studied the background from $Ta_2O_5$ at 532 nm excitation [10] and found a weak Raman background, which is located at short wavelengths. As the $Si_3N_4$-B had a stronger background than the $Si_3N_4$-S, we decided to only study the $Si_3N_4$-B further to obtain ab upper bound on the background emission. The $Si_3N_4$-B was further investigated by measuring the forward propagating light with a spectrometer. The intensity of the forward propagating background depends on the propagated distance, and more importantly, it varies with wavelength due to changes in propagation loss with wavelength. The coupled light will decay exponentially along propagation distance, but the background will have a maximum at the saturation distance $L_s=1/\alpha$, where α is the propagation loss. The measured spectrum thus has to be adjusted for the saturation length of the waveguide platform for the given wavelength.

The resulting spectra for 640 nm, 561 nm and 488 nm excitation is plotted in Figure 3a, with logarithmic axes. The spectra have been normalized in Fig. 3b, in order to more easily see the shape of the spectra. The spectrometer used in these experiments did not have high enough sensitivity to measure the weak background from $Ta_2O_5$, and since it did not have any effect on the imaging, it was not pursued further.

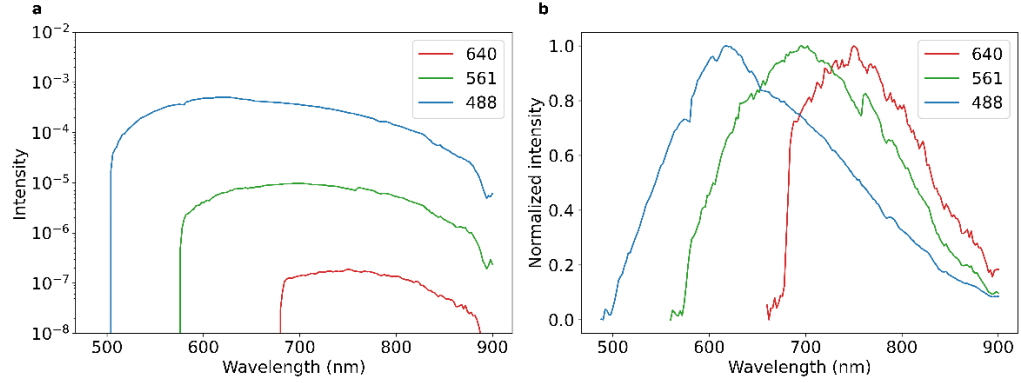

Fig. 3. (a) Background spectra for $Si_3N_4$-B for three different excitation wavelengths plotted on a logarithmic scale. (b) Normalized spectra.

*3.2 Effect of background on super-resolution imaging*

We investigated the background effect on single molecule blinking data fitting for $Ta_2O_5$, which exhibited the lowest background, and $Si_3N_4$-B, which exhibited the highest background. Each platform was imaged with 640 nm and 488 nm, as the background increases with shorter wavelengths. Example images of sparse blinks for each combination is presented in Figure 4, panels a-d. The localization uncertainty is of great interest in single molecule localization microscopy, as it determines the resolution of the image. Box plots of the localization uncertainty for each sample imaged are presented in panels e-h. The mean and standard deviation of the localization uncertainty, background standard deviation and intensity for all measurements are found in panels i-k. For $Ta_2O_5$, the average localization uncertainty is 12.9 ± 1.8 nm for 640 nm excitation and 13.9 ± 1.9 nm for 488 nm excitation. For $Si_3N_4$-B, the average blinking uncertainty is 17.0 ± 2.6 nm for 640 nm excitation and 22.6 ± 3.9 nm for 488 nm excitation. There is a difference in all parameters, even for 640 nm where $Si_3N_4$ has almost no background, but this may be due to differences in platform such as confinement factor. It is important to note that the fluorophores used for 640 nm and 488 nm are not the same, and as such the difference in parameters with wavelength can be expected. The most interesting point to note is that the parameters are consistent for 488 nm and 640 nm for $Ta_2O_5$, whereas for $Si_3N_4$-B the parameters are affected by the increased background at 488 nm compared to 640 nm.

The effect of a homogeneous background on the localization precision can be calculated using the following equation, assuming a background limited case [43]:

$$\langle(\Delta x)^2\rangle = \frac{4\sqrt{\pi}s^3 b^2}{aN^2}$$

Where $\Delta x$ is the localization error, *s* the standard deviation of the PSF, *b* the background noise, *a* the pixel size and *N* the number of photons collected. The values for *a* and *s* for the chip-based system is 106.5 nm and 113 nm, respectively. The number of photons and the background will depend on excitation wavelength for a background from the waveguide. Estimates of the background and photon counts for beads can be taken from Table 1 and converted into photons from counts. Typical values of AF647 blinks with the chip-based platform is around 500-1000 photons, which results in a localization precision as low as 15 nm for $Ta_2O_5$ and 16 nm for $Si_3N_4$-B. These match closely with the experimental results shown in Fig. 4i. For AF488 typical blink intensity on chips is around 260 photons, which gives a resolution of 62 nm in $Ta_2O_5$. The higher background in $Si_3N_4$-B will result in a loss in localization precision at 488 nm, giving 115 nm resolution for the high background case. This is not in agreement with the experimental results, but during the reconstruction process poor fits are rejected automatically, and as such the localization precision from the experiments will be artificially inflated. This means that even though the signal-to-background ratio on average is poor, data can still be reconstructed from the high intensity blinks. However, due to rejection of poorer blinking images, it could require a larger number of images during bio-imaging experiments.

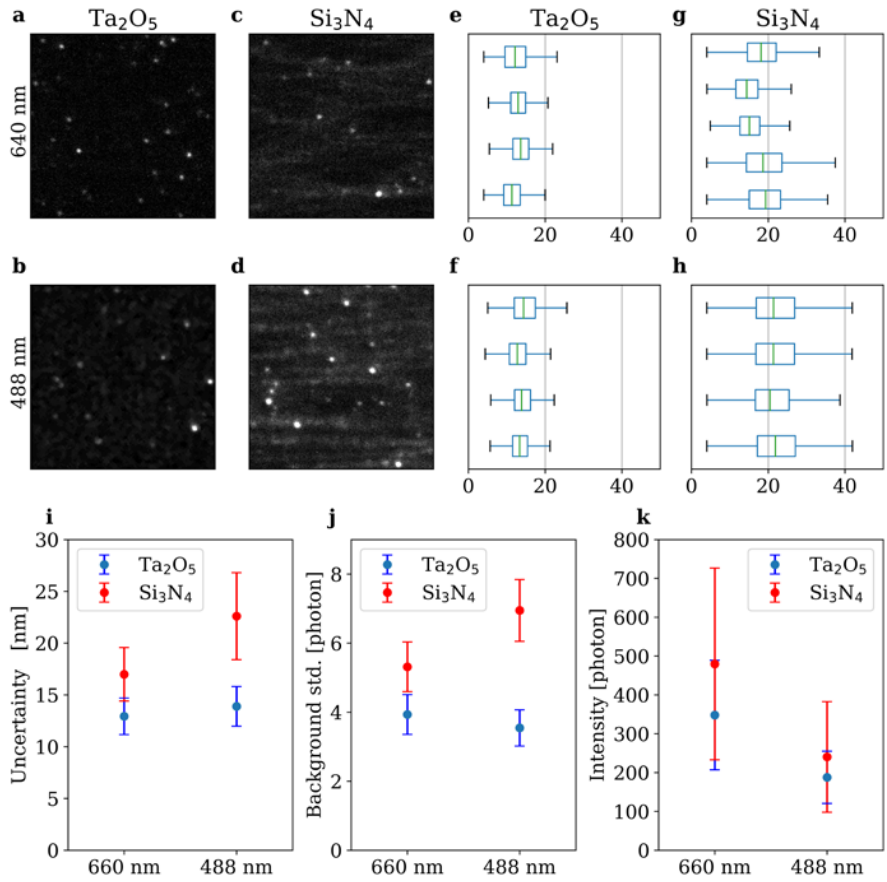

Fig. 4. (a-b) Representative image from *d*STORM data for $Ta_2O_5$ with AF647 and AF488. The imaged area is approximately 12 μm x 12 μm. (c-d) Representative image from *d*STORM data for $Si_3N_4$-B with AF647 and AF488. (e-f) Blinking statistics of reconstructed datasets for $Ta_2O_5$ with AF647 and AF488. Each box plot is a separate measurement; (g-h) Blinking statistics of reconstructed datasets for $Si_3N_4$-B with AF647 and AF488 image. The horizontal axis is in nanometers for panel (e-h). (i-k) Summary of localization uncertainty, background standard deviation and intensity for all measured samples.

The illumination generated by the optical waveguide is not uniform, thus the background and excitation distribution are not homogeneous, as assumed in the above calculation. There is instead a distinct mode pattern as can be observed in Fig. 1b. The waveguides used in the work are highly multi-mode and will therefore exhibit patterns such as that in Fig 1b. The specific pattern will depend on the coupling into the waveguides, but due to the structure of the waveguide they will all have a directionality along the propagation direction. Such an inhomogeneous background can introduce a difference in localization precision along the propagation direction compared to the perpendicular direction. Analysis of simulated data with experimentally captured background patterns showed that the localization uncertainty was 18.0 nm in the vertical direction and 13.7 in the horizontal direction for the low intensity blinks (200-500 photons). In the high intensity sample (1200-1500 photons), the results were 9.2 nm in the vertical direction and 6.2 in the horizontal direction. This means that there is a small degree of bias introduced by the directionality and inhomogeneity of the pattern. A stronger background will increase this effect, whereas for low background the difference will disappear. Testing it with $Ta_2O_5$ background showed that there was no difference in the uncertainties.

The final experiment was to investigate if adding a 488 nm laser when imaging with a 640 nm fluorescent dye would deteriorate the localization precision. A short wavelength laser is commonly used in single molecule localization microscopy for tuning of the on and off states during the imaging process. Imaging can take a while in *d*STORM, which results in fluorescent dye molecules stuck in dark states. A short wavelength laser can help relax the dye molecules in dark states back to a fluorescing state. A 488 nm excitation does give a much stronger background compared to 640 in our $Si_3N_4$ waveguides, so it is of interest to determine whether a 5-10% 488 nm power for reactivating the fluorophores will affect the localization precision. Reconstructions of the same sample with and without 488 nm excitation (in addition to the main 640 nm excitation) showed no detectable difference in the results for both silicon nitride and tantalum pentoxide waveguide materials.

## 4. Discussion and conclusion

We have studied the background from two commonly used waveguide materials, $Si_3N_4$ and $Ta_2O_5$, for applications in super-resolution optical microscopy and Raman spectroscopy in the visible part of the spectrum. $Ta_2O_5$ has very little background for all the tested excitation wavelengths, which agrees well with previous literature [10]. $Si_3N_4$, on the other hand, exhibits a relatively strong background at shorter wavelengths (488 nm). Previous literature on $Si_3N_4$ has reported the comparable performance of $Si_3N_4$ with respect to the background of $Ta_2O_5$ when pumped at 785 nm [44]. In agreement with the previous literature, we also observed comparable background between the two materials when excited at 640 nm, however, the background for $Si_3N_4$ significantly increases at shorter wavelengths. Both platforms are therefore well suited for *d*STORM imaging at 640 nm, whereas $Ta_2O_5$ is expected to perform better at shorter wavelengths. For shorter wavelengths, such as at 488 nm excitation, the background for $Si_3N_4$ reduces the localization precision for *d*STORM imaging.

Using fluorescent beads and exciting them with the evanescent field is a simple way to gauge the background of the waveguide for imaging purposes. There was a clear increase in background for $Si_3N_4$ at shorter wavelengths for both fabrication methods chosen in this study. The fact that the absorption and quantum yield of the beads vary with wavelength makes a direct comparison impossible for one material between different excitation wavelengths, as a change in signal-to-background could be due to a change in e.g. quantum yield of the beads with wavelength. A fall in signal-to-background at shorter wavelengths could also be due to a fall in quantum yield of the beads at shorter wavelengths, rather than an actual increase in the background. However, the beads should exhibit a similar behavior on all platforms over the different excitation wavelengths, although differences in e.g. confinement factor between the platforms may cause some deviations. We are not able to rule out any differences in e.g. confinement factor, but with the results from Figure 3, it is clear that increase in background between the platforms is mostly due to an increase in the background. Since the change is not the same for $Si_3N_4$ and $Ta_2O_5$, it means that the background in $Si_3N_4$ does increase at shorter wavelengths. This is further proven from the spectral measurements, as the background strength increases significantly at shorter wavelengths for $Si_3N_4$. From the spectroscopy experiment, it was evident that the background is a very wide fluorescent signal and thus it is not possible to filter it out very efficiently. The background originating in the waveguide, when irradiating at 488 nm, may have different origins (defects, composition, impurities in the silicon nitride core layers), but it seems that the stoichiometry shift could increase the UV absorption showing that films enriched with silicon show higher luminescence in the visible region of the electromagnetic spectrum [45].

The effect of the background signal is important to fluorescence optical microscopy. For diffraction limited imaging the background does reduce the image quality, but is not detrimental, as the signal-to-background ratio in fluorescence microscopy is usually good. For *d*STORM, where signals are observed from single molecules, the background clearly has an adverse effect on the result. The measured increase in e.g. localization uncertainty shown in Figure 4 is not drastic, but it is underestimated in the present experiment, where we have only localized the fluorophores placed directly on top of waveguide surface. According to the calculations on a homogeneous background from Equation 1, a significant loss in localization precision is expected for $Si_3N_4$ at 488 nm excitation. The reason that the experimental data underestimates the effect is that the reconstruction algorithm disregards localizations outside a given threshold. Thus, several blinks will be disregarded in the experimental data which could necessitate acquisitions of a larger number of images. It could thus be possible to perform *d*STORM at 488 nm, but it will both degrade the result and put restraints on data collection. Additionally, the inhomogeneous background pattern does cause a small bias in the localization precision. The bias has a maximum 4.3 nm in the low intensity case and 3.1 nm in the high intensity case. For $Ta_2O_5$ no such bias will occur, as there is no detectable background.

When imaging with 640 nm excitation and adding a small percentage (10%) of 488 nm excitation (for *d*STORM), it barely has any detectable difference. This was further investigated by simulations using the same background pattern from $Si_3N_4$ at 488 nm as above, but increasing the photon count tenfold, showed no discernable difference in vertical and horizontal root mean square error.

In summary, we have investigated $Si_3N_4$ and $Ta_2O_5$ for visible wavelength applications, in particular super-resolution optical microscopy and spectroscopic applications. $Si_3N_4$ has an increase in background at shorter wavelengths, whereas $Ta_2O_5$ has very weak background, effectively undetectable for imaging purposes. Despite the background of $Si_3N_4$ it might still be used for diffraction limited imaging. For super-resolution imaging at 488nm with SMLM, due to higher background with $Si_3N_4$, greater number of single molecule blinks rejected. It is also important to note the possibility of a bias due to the background pattern. With the increasing interest in integrated photonics employing visible light, further investigations into the origin of the background in $Si_3N_4$ at shorter wavelengths would be interesting to explore in the future, to determine if there are a correlation between the film composition and the background appearing.

**Funding.** B.S.A acknowledge the funding from UiT Tematiske Satsinger and Research Council of Norway (project #BIOTEK 2021-285571).

**Disclosures.** B.S.A. has applied for two patents for chip-based optical nanoscopy. The authors declare no conflicts of interest. B.S.A. and Ø.I.H are co-founders of the company Chip NanoImaging AS, which commercializes on-chip super-resolution microscopy systems.

**Data availability.** Data underlying the results presented in this paper are not publicly available at this time but may be obtained from the authors upon reasonable request.